\documentclass[11pt]{article}

\usepackage{fullpage}
\usepackage{latexsym}
\usepackage{amssymb,amsfonts,amsmath,amsthm}

\def\01{\{0,1\}}
\newcommand{\ceil}[1]{\lceil{#1}\rceil}

\newcommand{\eps}{\varepsilon}

\newcommand{\poly}{\mbox{\rm poly}}
\newcommand{\Exp}{{\mathbb{E}}}
\newcommand{\ket}[1]{|#1\rangle}

\newcommand{\Tr}{\mbox{\rm Tr}}

\renewcommand{\Pr}{\mbox{\rm Pr}}

\newtheorem{theorem}{Theorem}

\newtheorem{corollary}[theorem]{Corollary}

\renewcommand{\qed}{\hfill{\rule{2mm}{2mm}}}
\renewenvironment{proof}[1][]{\begin{trivlist}
\item[\hspace{\labelsep}{\bf\noindent Proof#1:\/}] }{\qed\end{trivlist}}

\begin{document}
\title{\bf Simultaneous Communication Protocols with Quantum and Classical Messages}
\author{Dmitry Gavinsky\thanks{NEC Laboratories America, Inc.
4 Independence Way, Suite 200. Princeton, U.S.A.}
\and Oded Regev\thanks{School of Computer Science, Tel-Aviv University, Tel-Aviv 69978, Israel. Supported
   by the Binational Science Foundation, by the Israel Science Foundation,
   by the European Commission under the Integrated Project QAP funded by the IST directorate as Contract Number 015848, and by a European Research Council (ERC) Starting Grant.
   }
 \and Ronald de Wolf\thanks{CWI, Kruislaan 413, 1098SJ Amsterdam, The Netherlands. Partially supported by Veni and Vidi grants from the Netherlands Organization for Scientific Research (NWO), and by the European Commission under the Integrated Project Qubit Applications (QAP) funded by the IST directorate as Contract Number 015848.}}
\maketitle

\begin{abstract}
We study the simultaneous message passing (SMP)  model of communication complexity, for
the case where one party is quantum and the other is classical.
We show that in an SMP protocol that computes some function with the first party sending $q$ qubits and the second
sending $c$ classical bits, the quantum message can be replaced by a \emph{randomized} message of
$O(qc)$ classical bits, as well as by a \emph{deterministic} message of $O(q c \log q)$ classical bits.
Our proofs rely heavily on earlier results due to Scott Aaronson~\cite{aaronson:advicecommj,aaronson:qlearnability}.

In particular, our results imply that quantum-classical protocols need to send
$\Omega(\sqrt{n/\log n})$ bits/qubits to compute {\sc Equality} on $n$-bit strings, and hence are not significantly
better than classical-classical protocols (and are much worse than quantum-quantum protocols
such as quantum fingerprinting).
This essentially answers a recent question of Wim van Dam~\cite{dam:commsmp}.
Our results also imply, more generally, that there are no superpolynomial separations between quantum-classical
and classical-classical SMP protocols for functional problems.
This contrasts with the situation for \emph{relational} problems,
where exponential gaps between quantum-classical and classical-classical SMP protocols are known.
We show that this surprising situation cannot arise in purely classical models: there, an exponential separation
for a relational problem can be converted into an exponential separation for a functional problem.
\end{abstract}

\section{Introduction}

We consider the simultaneous message passing (SMP) model of communication complexity.
Here Alice receives input $x$ and Bob receives input $y$. They each send one message to
a third party, called the ``referee.''  Given the two messages, the referee outputs
a value which should equal the function value $f(x,y)$ with probability at least, say, $2/3$.

We are interested in comparing classical and quantum SMP protocols.
Consider for instance the {\sc Equality} function:  $x,y\in\01^n$, and $f(x,y)=1$ iff $x=y$.
If Alice and Bob do not share randomness, then this function exhibits an exponential
quantum-classical gap: there is an SMP protocol for $f$ where Alice and Bob
each send $O(\log n)$ \emph{quantum} bits to the referee~\cite{bcww:fp}, using
a technique called ``quantum fingerprinting.''
On the other hand, if the messages are classical, then $\Theta(\sqrt{n})$-bit messages are necessary and
sufficient~\cite{ambainis:3computer,newman&szegedy:1round,babai&kimmel:simultaneous},
as we will explain in Section~\ref{secwarmup}.

Here we consider a question recently asked by Wim van Dam~\cite{dam:commsmp}:
what happens if one of the messages (say Alice's) is quantum, while the other is restricted to be classical?
We call such protocols \emph{quantum-classical} SMP protocols.
For instance, one may ask whether some variant of quantum fingerprinting still works
if one of the two messages is classical.

Our main results here say that the quantum message can be ``simulated'' by a classical
message that is only moderately larger. More specifically, we show that the following hold
for protocols computing a Boolean function:
\begin{itemize}
\item
In Theorem~\ref{threplacequantum} we show that a private-coin (resp.~public-coin) quantum-classical
protocol where Alice sends $q$ qubits and Bob sends $c$ bits,
gives a private-coin (resp.~public-coin) classical-classical protocol where Alice sends a randomized $O(qc)$-bit
message and Bob sends a randomized $c$-bit message.
\item
In Corollary~\ref{cordeterministicsimsmp} we show that a private-coin quantum-classical protocol
where Alice sends $q$ qubits and Bob sends $c$ bits,
gives a private-coin classical-classical protocol where Alice sends a \emph{deterministic} $O(qc \log q)$-bit message
and Bob sends a randomized $c$-bit message.
\end{itemize}
Our proofs rely heavily on earlier results in the one-way communication model
by Aaronson~\cite{aaronson:advicecommj,aaronson:qlearnability}.

The latter result implies that quantum-classical private-coin protocols for {\sc Equality}
need to send $\Omega(\sqrt{n/\log n})$ bits and/or qubits.  This is not significantly
better than the $\Theta(\sqrt{n})$ bits that are necessary and sufficient
for classical protocols.

The results above are rather unusual in communication complexity, where typically
introducing quantum elements gives exponentially more power for at least some problems.
The exponential improvements given by the quantum-quantum fingerprinting protocol is a prime example of this.
This is also the case with one-way communication complexity,
where we know Boolean functions whose quantum complexity is exponentially
smaller than their classical complexity~\cite{gkkrw:1way}.
As another example, Gavinsky et al.~\cite{gkkrw:1way} exhibit
a function with an exponential gap between classical-classical SMP protocols with shared
entanglement and those with only shared randomness. Yet another example is
obtained when considering \emph{relations} instead of \emph{functions} in our model~\cite{bjk:q1way}.
We elaborate on this in Section~\ref{secfnrel}.


\section{Preliminaries}

We assume familiarity with quantum computing~\cite{nielsen&chuang:qc} and with the basic
notions of classical and quantum communication complexity~\cite{kushilevitz&nisan:cc,wolf:qccsurvey}.
Informally, the setting of communication complexity is as follows.
Alice receives some input $x\in X$, Bob receives some input $y\in Y$,
and together they want to compute some function $f(x,y)$, with error probability at most $1/3$ for
all $(x,y)$ in some domain ${\cal D}\subseteq X\times Y$.
If ${\cal D}=X\times Y$ then the problem is called \emph{total}, otherwise it is called \emph{partial}.
A \emph{Boolean} function $f$ has range $\01$.
The \emph{communication matrix} $M_f$ corresponding to $f$ is the $|X|\times |Y|$ matrix whose
$(x,y)$-entry equals $f(x,y)$ if $(x,y)\in{\cal D}$, and equals `*' otherwise.

In the simultaneous message passing (\emph{SMP}) model, Alice and Bob each send a message to a
third party (called the \emph{referee}), who then computes the output.
In the \emph{one-way} model Alice sends one message to Bob
(no referee here), while in the \emph{two-way} model they can interact arbitrarily.
The \emph{cost} of a communication protocol is its total communication on the worst-case input.
The (bounded-error) communication complexity of $f$ (in one of the above models) is the minimal cost among all protocols
that compute $f$ with probability of error at most $1/3$ for each input.

Classical protocols may be deterministic or randomized.
When dealing with randomness, we have to distinguish between \emph{private coins}
and \emph{public coins}.\footnote{We will not consider models with shared entanglement here.}
The former are visible only to individual parties, while the latter
are shared among all parties and may help them coordinate their actions.
In one-way and two-way models, the difference between public-coin and private-coin
communication complexity is at most an additive $O(\log n)$ bits~\cite{newman:random},
but in the SMP model it can make a big difference.  In particular, while the private-coin SMP
complexity of {\sc Equality} is $\Theta(\sqrt{n})$ bits, its public-coin complexity is constant:
the protocol picks a shared random $n$-bit string $r$, Alice and Bob send the inner product
of their input with $r$ (mod 2) to the referee, who checks whether the two received bits are equal.

\smallskip
The following theorem can be derived from the proof of~\cite[Theorem~1.4]{aaronson:qlearnability}.
Here by a measurement operator $E$ we mean a positive semidefinite matrix with eigenvalues in $[0,1]$.
The \emph{acceptance probability} of the two-outcome measurement with operators $E$ and $I-E$
on density matrix $\rho$ is $p=\Tr(E\rho)$.

\begin{theorem}[Aaronson]\label{thlearningstate}
For all $\delta>0$ the following holds.
Suppose Alice has the classical description of an arbitrary $q$-qubit density matrix~$\rho$
and Bob has a measurement operator $E\in\{E_b\}_{b\in\01^c}$.
Then Alice can send Bob a randomized $O(qc)$-bit message that enables him to output a value $p'$ that
approximates $p=\Tr(E\rho)$ in the following sense: with probability at least $1-\delta$ we have $|p-p'|\leq \delta$
(the constant in the $O(\cdot)$ depends on $\delta$, and no public coin is used).
\end{theorem}

\section{Warmup: replacing a randomized message by deterministic}\label{secwarmup}

Our goal in this paper is to replace a quantum message by a randomized or deterministic one
that is not much bigger.  Let us first consider a known case:
replacing a \emph{randomized} message by a \emph{deterministic} one.
Babai and Kimmel~\cite{babai&kimmel:simultaneous} showed the following.
If there is a bounded-error private-coin SMP protocol for a Boolean function $f$,
where Alice sends $c_A$ bits and Bob sends $c_B$ bits, then there exists another
bounded-error private-coin SMP protocol for $f$ where Alice \emph{deterministically}
sends $O(c_A c_B)$ bits, and Bob (who is still randomized) sends $c_B$ bits.

As an application of this result, note that in any bounded-error protocol
for {\sc Equality} where Alice is deterministic,
she needs to send a different message for each of the $2^n$ inputs $x$:
if she sends the same message for $x$ and for $x'$, then the referee will have
the same acceptance probability on inputs $(x,x)$ and $(x',x)$ and hence on at
least one of those pairs he will err with probability at least 1/2.
This implies that in any bounded-error private-coin SMP protocol for {\sc Equality},
$n \le O(c_A c_B)$, and hence we obtain the bound $c_A+c_B \ge \Omega(\sqrt{n})$
mentioned in the introduction.

As a warm-up for the results in the next sections, we sketch a simple proof of the Babai-Kimmel result here.
Consider any Boolean function $f$, partial or total, and a bounded-error private-coin SMP protocol for $f$.
Let $r(a,b)\in[0,1]$ denote the referee's acceptance probability if he receives message
$a\in\01^{c_A}$ from Alice and $b\in\01^{c_B}$ from Bob. We use $A_x$ for the random variable which
is Alice's message (its distribution depends on her input $x$), and similarly $B_y$ for Bob's message.
Note that for every input $x,y$ where $f$ is defined, the acceptance probability
$\Exp_{a\sim A_x,b\sim B_y}[r(a,b)]$ approximates the function value:
$|f(x,y)-\Exp_{a\sim A_x,b\sim B_y}[r(a,b)]|\leq 1/3$.

We now modify the protocol as follows.
Let Alice send her (probabilistic) message $s\cdot c_B$ times, for some integer $s$
to be determined later, at a total communication cost of $s c_A c_B$ bits.
For a fixed message $b$ from Bob, the referee computes $\widetilde{p}_b$,
defined as the average of $r(a_i,b)$ over all messages $a_i$ received from Alice.
Choosing $s$ a sufficiently large constant, the Chernoff bound implies that $\widetilde{p}_b$
is within 1/10 of its expectation $p_b=\Exp_{a\sim A_x}[r(a,b)]$, except with probability $\ll 2^{-c_B}$:
$$
\Pr[|\widetilde{p}_b-p_b|>1/10]\ll 2^{-c_B}.
$$
Hence by a union bound, there exists a deterministic message from Alice that
would cause the approximations $\widetilde{p}_b$ to be $1/10$-close to $p_b$
simultaneously for all $b\in\01^{c_B}$. The desired protocol is clear now:
Alice, given her input $x$, sends the $s c_A c_B$-bit deterministic message
described above. The referee uses this information to compute a value
$\tilde{p}_b$ such that $|\widetilde{p}_b-p_b| \le 1/10$ where
$b$ is the message he received from Bob, and accepts with probability
$\tilde{p}_b$. Clearly, for each input pair $x,y$ the acceptance
probability of this protocol differs from that of the original
protocol by at most $1/10$, and hence correctness is maintained.
We have reproved:

\begin{theorem}[Babai \&\ Kimmel]\label{thbabaikimmel}
Let $f$ be a (possibly partial) Boolean function.
If there is a bounded-error private-coin SMP protocol for $f$ where Alice sends $c_A$ bits
and Bob sends $c_B$ bits, then there is a bounded-error private-coin SMP protocol for $f$ where Alice
\emph{deterministically} sends $O(c_A c_B)$ bits, and Bob (who is still randomized) sends $c_B$ bits.
\end{theorem}

Actually, Babai and Kimmel showed something slightly stronger, namely that both Alice's and
Bob's randomized message can simultaneously be replaced by deterministic messages of length $O(c_A c_B)$.

\begin{theorem}[Babai \&\ Kimmel]
Let $f$ be a (possibly partial) Boolean function.
If there is a bounded-error private-coin SMP protocol for $f$ where Alice sends $c_A$ bits
and Bob sends $c_B$ bits, then there is a deterministic SMP protocol for $f$ where Alice and Bob
send $O(c_A c_B)$ bits.
\end{theorem}

This theorem is essentially tight for the {\sc Equality} problem,
where the deterministic communication complexity is $2n$.
Consider the following bounded-error private-coin protocol, adapted from~\cite{ambainis:3computer}.
Alice and Bob fix, beforehand, a good error-correcting code $C:\01^n\rightarrow\01^m$ with $m=O(n)$.
For $m=c_A c_B$, they can view the codewords as $c_A\times c_B$ Boolean matrices.
Alice sends a random column of $C(x)$ together with its index, sending $c_A+\log c_B$ bits in total.
Bob sends a random row of $C(y)$ with its index, sending $c_B+\log c_A$ bits.
The referee checks whether the row and column agree in the one point where they intersect.
If $x=y$ then the two bits are the same, otherwise they differ with constant probability.
Repeating a constant number of times, we obtain a bounded-error private-coin SMP protocol where Alice
sends $O(c_A+\log c_B)$ bits and Bob sends $O(c_B+\log c_A)$ bits.

Note that Theorem~\ref{thbabaikimmel} fails spectacularly for \emph{public-coin} SMP protocols:
the bounded-error public-coin SMP complexity of {\sc Equality} is constant, while a deterministic player
needs to send $n$ bits, no matter what the other player sends.

\section{Replacing a quantum message by a randomized one}

We now prove that in quantum-classical SMP protocols, the ``quantum leg'' of the protocol can be replaced by a randomized message.

\begin{theorem}\label{threplacequantum}
Let $f$ be a (possibly partial) Boolean function.
If there is a private-coin (resp.~public-coin) bounded-error quantum-classical SMP protocol for $f$
where Alice sends $q_A$ qubits and Bob sends $c_B$ classical bits,
then there is a private-coin (resp.~public-coin) bounded-error SMP protocol for $f$ where Alice
sends $O(q_A c_B)$ classical bits and Bob sends $c_B$ classical bits.
\end{theorem}

\begin{proof}
We prove the theorem for private-coin protocols.  The proof for public-coin protocols is essentially
the same, just fix the shared randomness at the start of the argument and average over it at the end.
In general the three-party protocol has the following form: Alice sends the referee a $q_A$-qubit
density matrix $\rho_x$, while Bob sends a classical message $b\in\01^{c_B}$, whose distribution depends
on his input $y$. The referee then measures $\rho_x$ with a measurement operator $E_b$,
and outputs 1 if the measurement accepts (which happens with probability $p_b=\Tr(E_b\rho_x)$).

The SMP protocol promised by the theorem is as follows:
Bob sends a $c_B$-bit message $b$, exactly as in the original quantum-classical protocol.
Using Theorem~\ref{thlearningstate},
Alice sends to the referee a randomized message of $O(q_A c_B)$ bits to enable him to obtain with probability at least $1-\delta$
an approximation $\widetilde{p}_b$ to $p_b=\Tr(E_b\rho_x)$ to within $\pm \delta$, where $\delta=1/10$.
Finally, the referee outputs $1$ with probability $\widetilde{p}_b$, and 0 otherwise.
The overall error probability is at most $2\delta$ worse than in the original protocol.
\end{proof}


Note the difference between the above two proofs. The proof of Theorem~\ref{thbabaikimmel}
obtains approximations $p_b$ for all $b\in\01^{c_B}$ simultaneously, which enables the referee
to learn $f(x,y)$ for each $y$.
On the other hand, the proof of Theorem~\ref{threplacequantum} only obtains an approximation $p_b$
for the specific $b$ that the referee received from Bob, which enables the referee to predict
$f(x,y)$ for the specific input $y$ that Bob holds.

\section{Replacing a quantum message by a deterministic one}

By combining Theorem~\ref{threplacequantum} with the
Babai-Kimmel theorem (Theorem~\ref{thbabaikimmel}), we see that
in every quantum-classical private-coin SMP protocol with $q_A$ qubits and $c_B$ bits,
we can replace Alice's message by a \emph{deterministic} message of $O(q_A c_B^2)$ bits.
However, we can obtain something that is usually stronger,
namely a deterministic message of $O(q_A c_B \log q_A)$ bits.
The crucial tool is the following result, which is an extension of a result of
Aaronson~\cite[Theorem~3.4]{aaronson:advicecommj}:

\begin{theorem}\label{thdeterministicsim}
For all $\delta>0$ the following holds.
Suppose Alice has the classical description of an arbitrary $q$-qubit density matrix~$\rho$,
and Bob has $2^c$ measurements operators $\{E_b\}_{b\in\01^c}$.
There is a deterministic message of $O(q c \log q)$ bits from Alice
that enables Bob to output values $p'_b$ satisfying that $|p_b-p'_b|\leq \delta$ simultaneously for all $b\in\01^c$
where $p_b=\Tr(E_b\rho)$.
\end{theorem}

It is interesting to compare this with Theorem~\ref{thlearningstate}.
While Theorem~\ref{thlearningstate} allows us to approximate one $p_b$ to within $\pm \delta$ (with some small probability of error)
using an $O(q c)$-bit message, Theorem~\ref{thdeterministicsim} allows us to approximate \emph{all} $p_b$
to within $\pm \delta$ (\emph{without} probability of error) at the expense of increasing the message length by a factor $\log q$.
Theorem~\ref{thdeterministicsim} generalizes Aaronson's~\cite[Theorem~3.4]{aaronson:advicecommj},
which proves the special case where $\Tr(E_b\rho)$ is close to 0 or 1 for all $b$.
Our proof is a modification of his. We conjecture that the $\log q$ factor is not needed.

\begin{proof}
Suppose Alice sends $r=O(\log q)$ many copies of her state, for sufficiently large constant in the $O(\cdot)$ which may depend on $\delta$.
Let $\rho'=\rho^{\otimes r}$
be the state she sends, and $K=r q=O(q \log q)$ its total number of qubits.
Define the observable\footnote{An \emph{observable} $F$ is a Hermitian matrix that describes a measurement,
as follows. By diagonalization we can write $F=\sum_i\lambda_i P_i$, where $P_i$ is the projector on
the eigenspace corresponding to eigenvalue $\lambda_i$. These eigenspaces are all orthogonal to each other and $\sum_i P_i = I$.
The corresponding measurement on a density matrix $\rho$ gives outcome $\lambda_i$ with probability $\Tr(P_i\rho)$.
Hence the \emph{expectation} of the measurement is $\sum_i\lambda_i\Tr(P_i\rho)=\Tr(F\rho)$.}
$$
F_b=\frac{1}{r}\sum_{j=1}^r E_b^{(j)},
$$
where $E_b^{(j)}$ applies $E_b$ to the $j$th copy.
This measures the fraction of successes if you separately measure each of the $r$ copies of $\rho$ with $E_b$.
Since $r=O(\log q)$, a Chernoff bound implies that the outcome $p_b'$ of this measurement applied to $\rho'$ will probably be close
to its expectation $p_b=\Tr(E_b\rho)$:
\begin{equation}\label{equsechernoff}
\Pr[|p'_b-p_b|>\delta/4]\leq 1/\poly(q).
\end{equation}

{\bf Alice's classical message.}
Consider all $b=1,\ldots,2^c$ in order.
We will sequentially build a sequence of $K$-qubit density matrices $\rho_b$, one for each $E_b$.
Alice's classical message will enable Bob to reconstruct this entire sequence.
Call $b$ \emph{good} if $|\Tr(F_b\rho_b)-p_b|\leq\delta$; call $b$ \emph{bad} otherwise.
Note that if Bob has a classical description of a good $\rho_b$, then he can approximate $p_b$ to within $\pm \delta$
(since he knows what $F_b$ is).
We start with the completely mixed state: $\rho_1=I/2^K$ and define the subsequent $\rho_b$ one by one, as follows.
If $b$ is good, then define $\rho_{b+1}$ to be equal to $\rho_b$.
If $b$ is bad, Alice appends the pair $(b,\widetilde{p}_b)$ to her message, where $\widetilde{p}_b$
is the $\log(1/\delta)+O(1)$ most significant bits of $p_b$, so $|\widetilde{p}_b-p_b|\ll \delta$.
In this case, let $M_b$ be the projector on the subspace spanned by the eigenvectors of $F_b$
with eigenvalues in the interval $[\widetilde{p}_b-\delta/2,\widetilde{p}_b+\delta/2]$,
and let $\rho_{b+1}$ be the renormalized projection of $\rho_b$ on this subspace.\footnote{The fact that
this projection is nonzero (and hence can be renormalized to have trace~1)
follows from the argument in the ``Second, the lower bound on $p$'' paragraph below.}
Continuing all the way to $b=2^c$,
we obtain a message $(b_1,\widetilde{p}_{b_1}),\ldots,(b_T,\widetilde{p}_{b_T})$ for some $T$.
We need to show two things:
(1) this message enables Bob to approximate all $p_b$ to within $\pm \delta$,
and (2) $T=O(K)$, which implies that the message length is $O(q c \log q)$ bits.
We will show these two things in turn.

{\bf (1) Why this works.}
Note that Bob knows which $b\in[2^c]$ are bad, since those $b$ are exactly the ones in Alice's message.
Bob can in fact compute the whole sequence $\rho_1,\ldots,\rho_{2^c}$ given the message:
$\rho_1=I/2^K$; if $b$ is good then $\rho_{b+1}=\rho_b$; if $b$ is bad then
$(b,\widetilde{p}_b)$ is part of Alice's message and $\rho_{b+1}$ can be computed from this information.
Suppose Bob wants to approximate $p_b=\Tr(E_b\rho)$. If $b$ is good then by definition
$|\Tr(F_b\rho_b)-p_b|\leq\delta$ and Bob can calculate $\Tr(F_b\rho_b)$.
If $b$ is bad, then the pair $(b,\widetilde{p}_b)$ is part of Alice's message, so Bob knows $p_b$ with sufficient precision.
Hence Bob can approximate all $p_b$ up to $\pm \delta$, for all $b$ simultaneously.

{\bf (2) Why the message is not too long.}
Here we show $T=O(K)$. Define $\eta=1-\delta/4$ and $t=\ceil{(K+1)/\log(1/\eta)+1}=O(K)$.
Suppose, by way of contradiction, that $T\geq t$.
We consider the sequence $b_1,\ldots,b_t$ of the first $t$ bad $b$'s.
Let
$$
p=\Tr\left( M_{b_t} \cdots M_{b_1} \frac{I}{2^K} M_{b_1} \cdots M_{b_t} \right)
$$
be the probability that all $t$ measurements succeed if we start with the completely mixed state
and sequentially measure $M_{b_1},\ldots,M_{b_t}$. We will derive contradicting upper and lower bounds on~$p$.

First, the upper bound on $p$.
If we sequentially measure $M_{b_1},\ldots,M_{b_t}$, starting from the completely mixed state,
and if all $t$ measurements succeed, then we exactly have the sequence of density matrices
$\rho_{b_1}=I/2^K,\ldots,\rho_{b_t},\rho_{b_t+1}$.
We claim that if $\rho_b$ is bad, then $\Tr(M_b\rho_b)\leq \eta$,
and hence the probability that all $t$ measurements succeed is $p\leq \eta^t$.
The claim follows easily from a Markov argument: Let $X$ denote the random variable
representing the outcome of measuring $\rho_b$ with
the observable $F_b$. Notice that $X$ takes values in $[0,1]$.
Assume $\Tr(M_b\rho_b)=\Pr[|X-\widetilde{p}_b|\leq\delta/2] > \eta$.
Then $\Tr(F_b\rho_b) = \Exp[X]$ must necessarily be in the range
$$[\eta (\widetilde{p}_b - \delta/2), \eta (\widetilde{p}_b + \delta/2) + 1-\eta] \subseteq
[\widetilde{p}_b - 3\delta/4, \widetilde{p}_b + 3\delta/4] \subseteq
[p_b - \delta, p_b + \delta],$$
and hence $\rho_b$ is good.

Second, the lower bound on $p$.
Note that $M_b$ succeeds on $\rho'$ iff the outcome $p'_b$ of the observable $F_b$ is at most $\delta/2$
away from the number $\widetilde{p}_b$, which is the truncated version of $p_b=\Tr(E_b\rho)$
(recall $|\widetilde{p}_b-p_b|\ll \delta$).
Hence by Eq.~\eqref{equsechernoff}, we have
$$
\Tr(M_b\rho')=\Pr[|p'_b-\widetilde{p}_b|\leq\delta/2]\geq \Pr[|p'_b-p_b|\leq\delta/4]\geq 1-1/\poly(q).
$$
This allows us to measure $\rho'$ with $M_b$ while disturbing the state by only an insignificant amount.%
\footnote{This is a well-known and very intuitive property in quantum computing;
we refrain from spelling out the quantitative details so as not to crowd out the intuition with even more technicalities
(see for instance the ``almost as good as new lemma''~\cite[Lemma~2.2]{aaronson:advicecommj}).}
If we measure each of $M_b$, for the first $t$ bad $b$'s in sequence, starting in $\rho'$,
then with probability at least 1/2 all measurements will succeed.
However, the completely mixed state can be written as
$\frac{I}{2^K}=\frac{1}{2^K}\rho'+(1-\frac{1}{2^K})\rho''$
where $\rho''$ is orthogonal to $\rho'$. Hence if we start from $I/2^K$,
then the probability of all measurements succeeding is $p\geq 1/2^{K+1}$.

Combining the bounds of the last two paragraphs together with our value of $t$ gives a contradiction.
\end{proof}

\begin{corollary}\label{cordeterministicsimsmp}
Let $f$ be a (possibly partial) Boolean function.
If there is a bounded-error quantum-classical private-coin SMP protocol for $f$ where Alice sends $q_A$ qubits
and Bob sends $c_B$ bits, then there is a bounded-error private-coin SMP protocol for $f$ where Alice
\emph{deterministically} sends $O(q_A c_B \log q_A)$ bits, and Bob (who is still randomized) sends $c_B$ bits.
\end{corollary}

\begin{proof}
As in the proof of Theorem~\ref{threplacequantum}, we assume without loss of generality
that the protocol has the following form: Alice sends the referee a $q_A$-qubit
density matrix $\rho_x$, while Bob sends a classical message $b\in\01^{c_B}$, whose distribution depends
on his input $y$. The referee then measures $\rho_x$ with a measurement operator $E_b$,
and outputs 1 if the measurement accepts (which happens with probability $p_b=\Tr(E_b\rho_x)$).

The desired SMP protocol is as follows:
Bob sends a $c_B$-bit message $b$, exactly as in the original quantum-classical protocol.
Using Theorem~\ref{thdeterministicsim},
Alice sends to the referee a deterministic message of $O(q_A c_B \log q_A)$ bits to enable him
to obtain (with certainty) an approximation $\widetilde{p}_b$ to $p_b=\Tr(E_b\rho_x)$
to within $\pm \delta$, where $\delta=1/10$.
Finally, the referee outputs $1$ with probability $\widetilde{p}_b$, and 0 otherwise.
The overall error probability is at most $\delta$ worse than in the original protocol.
\end{proof}

As argued in Section~\ref{secwarmup}, in a bounded-error protocol for {\sc Equality} where Alice is deterministic,
she needs to send at least $n$ bits.
Hence we obtain the following lower bound on quantum-classical private-coin SMP
protocols for {\sc Equality}, which is tight up to the $\sqrt{\log n}$-factor.

\begin{corollary}\label{corequality}
Every quantum-classical private-coin SMP protocol for {\sc Equality} has communication complexity $\Omega(\sqrt{n/\log n})$.
\end{corollary}

%

\section{Tightness}\label{sectightness}

The example of the {\sc Equality} function shows that Theorems~\ref{thdeterministicsim}
and Corollary~\ref{cordeterministicsimsmp} are essentially tight.

We do not know whether Theorem~\ref{threplacequantum} is close to optimal, but at least it shows
that the gap between quantum-classical and classical-classical SMP protocols is at most polynomial.
The following communication problem, adapted from~\cite{bjk:q1way,gkrw:identification,gkkrw:1way},
presents an interesting quantum-classical public-coin protocol that uses about $n^{1/3}$ qubits.
We do not know an equally efficient classical-classical public coin protocol for this problem;
the best one we know sends about $\sqrt{n}$ bits. This suggests that quantum-classical SMP protocols
can at least have \emph{some} polynomial advantage over classical-classical protocols.

The problem is as follows.
Let $n$ be an even integer.
Alice receives $x\in\01^n$, while Bob receives a perfect matching $M$
(i.e., a partition of $[n]=\{1,\ldots,n\}$ into $n/2$ disjoint pairs, called ``edges''), and a string $w\in\01^{n/2}$
whose bits are indexed by the edges in $M$.  We can view the edges of $M$ as rows (of weight 2)
in an $n/2\times n$ matrix $M$ over $GF(2)$. Then the matrix-vector product $Mx$
is the $n/2$-bit string obtained by taking, for each edge $(i,j)$ of $M$ in order,
the XOR $x_i\oplus x_j$.  The promise is that the Hamming distance between $w$ and $Mx$
is either at most $n/6$ or at least $n/3$,
and the function value is 1 in the first case and 0 in the second.

One can show easily that the deterministic complexity of the problem is $\Omega(n)$, as follows.
By the probabilistic method, there exists a set $S\subseteq\01^n$ of size $|S|=2^{\Omega(n)}$
such that all distinct $x,x'\in S$ are at distance around $n/2$. But then for each distinct
$x,x'\in S$ we can find a matching $M$ where the $n/2$-bit strings $Mx$ and $Mx'$ have distance close to $n/2$
(pick as many edges as possible that have one endpoint in a bitlocation where $x$ and $x'$ agree,
and one endpoint where they disagree). Putting $w=Mx$, we have $f(x,M,w)=1$ but $f(x',M,w)=0$.
Hence in a deterministic protocol, Alice will need to send a different message for each of the $x\in S$.
Therefore the deterministic SMP (and even one-way) communication complexity of this function is $\Omega(n)$.

On the other hand, here is a bounded-error public-coin SMP protocol where Alice sends $q_A\approx n^{1/3}$
qubits and Bob sends $c_B\approx n^{1/3}$ bits.
Alice and Bob use the public coin to select a random subset $S\subseteq[n]$ of about $n^{2/3}$ elements from $[n]$.
Now with high probability, $M\cap(S\times S)$ will contain $\Theta(n^{1/3})$ edges.
Alice sends the referee $n^{1/3}$ copies of the uniform superposition
$\frac{1}{\sqrt{|S|}}\sum_{i\in S}(-1)^{x_i}\ket{i}$.
Bob sends over $\Theta(n^{1/3})$ edges in $M\cap(S\times S)$, together with the corresponding bits of $w$.
The referee constructs two-dimensional measurement operators from the edges he received from Bob,
and measures each of the quantum states with them.  With probability close to~1, one of those measurements
will succeed and give him a bit of the string $Mx$.
Since the location of that bit in $Mx$ is random, comparing that bit with the corresponding
bit of $w$ (which is part of Bob's message), gives the referee the function value with probability at least 2/3.%
\footnote{The best classical protocol that we know works similarly.  It selects a set $S$ of about $\sqrt{n}$ elements,
and Alice sends the corresponding bits of $x$ to the referee at the expense of about $\sqrt{n}$ bits of communication.}

\section{Functional separations versus relational separations}\label{secfnrel}

As mentioned before, Theorem~\ref{threplacequantum} implies that the gap between quantum-classical
and classical-classical SMP protocols is at most polynomial for any Boolean function.
In contrast, Bar-Yossef et al.~\cite{bjk:q1way} exhibited a \emph{relational} problem
for which quantum-classical SMP protocols are \emph{exponentially} better than classical-classical
SMP protocols with a public coin.
In a relational problem, for each input pair $(x,y)$ there is a \emph{set} of valid outputs $z$.
In the case of~\cite{bjk:q1way}, Alice receives an arbitrary string $x\in\01^n$ and Bob
receives a perfect matching $M$ on $[n]$ from a set of $\Theta(n)$ possible perfect matchings.
Their goal is to output any $z$ of the form $(i,j,x_i\oplus x_j)$ where $(i,j)\in M$.

Bar-Yossef et al.\ exhibit a quantum-classical SMP protocol that solves this problem with success probability~1,
with a $\log n$-qubit message from Alice to the referee and a $\log n$-bit message from Bob.
Their protocol works as follows: Alice sends the referee a uniform superposition
$$
\frac{1}{\sqrt{n}}\sum_{i=1}^n(-1)^{x_i}\ket{i}.
$$
Bob sends the referee the index of his matching $M$ in the set of $\Theta(n)$ possible matchings,
which takes $\log n+O(1)$ classical bits.
Viewing $M$ as a measurement consisting of $n/2$ orthogonal two-dimensional projectors (one for each edge $(i,j)\in M$),
the referee measures the quantum state he received from Alice.  This gives him the state
$$
\frac{1}{\sqrt{2}}\left((-1)^{x_i}\ket{i}+(-1)^{x_j}\ket{j}\right)=\frac{(-1)^{x_i}}{\sqrt{2}}\left(\ket{i}+(-1)^{x_i\oplus x_j}\ket{j}\right)
$$
for a uniformly random edge $(i,j)\in M$.
{}From this the referee can obtain the bit $x_i\oplus x_j$ with certainty, and output $(i,j,x_i\oplus x_j)$ accordingly.
No public coin is needed. In contrast, Bar-Yossef et al.\ proved a $\Theta(\sqrt{n})$ bound
for classical-classical public-coin protocols for this relational problem.

To summarize, we see that when comparing the quantum-classical SMP model to the classical-classical SMP model, one can obtain an exponential separation for a relation but not for a Boolean function. In the remainder of this section we
show that such a situation \emph{cannot occur} for purely classical models. More precisely, we show that for models obeying the \emph{Yao principle}~\cite{yao:unified} (defined next), any separation for a relation implies a similar separation for a function (and sometimes even for a Boolean function). Notice that the converse implications clearly holds, i.e., a separation for a functional problem is also a separation for a relational problem, since functions are a special case of relations.

Consider any computational model that has a class
of deterministic algorithms (or protocols), each of a certain cost.
A \emph{randomized} algorithm in such a model is a probability distribution over deterministic algorithms.
The Yao principle states the equality of two different complexities:
(1) the $\eps$-error complexity (the minimal cost of randomized algorithms
whose error probability is at most $\eps$ on every input) and (2) the $\eps$-error distributional complexity under
the hardest input distribution $\mu$  (the minimal cost of deterministic
protocols that have error probability at most $\eps$ under $\mu$).
Because of the minimax theorem from game theory, allowing shared randomness
in one's communication model is a sufficient condition for the Yao principle to hold.

Now assume any two computational models, both obeying the Yao principle.
Call these models ``a'' and ``b'', respectively.
Let $R_{a,\eps}$ and $R_{b,\eps}$ denote the $\eps$-error complexities
in these two models. Assume we have a separation between these two models, showing that
$R_{a,2\eps}(P)$ is greater than $R_{b,\eps}(P)$ for some relational problem~$P$. We will next show how to construct a function $f$ for which there is a separation between $R_{a,\eps}(f)$ and $R_{b,\eps}(f)$ that is at least as large.

Let $\mu$ be a worst-case input distribution for the distributional $R_{a,2\eps}$-complexity of relation $P$. That is, any deterministic protocol in the ``a'' model solving $P$ with distributional error at most $2\eps$ under $\mu$, must use at least $R_{a,2\eps}(P)$ bits of communication. Next, by the Yao principle, there is a deterministic protocol in the ``b'' model solving $P$ with error at most $\eps$ under distribution $\mu$, using at most $R_{b,\eps}(P)$ bits of communication. Being deterministic, this protocol necessarily computes some function with error probability 0, call it~$f$. Hence we see that for this function, $R_{b,\eps}(P) \ge R_{b,0}(f)$. Moreover, note that the probability under $\mu$ that $f(x,y)$ is not a valid answer for $P$, is at most $\eps$.

To complete the argument, we now show that $R_{a,\eps}(f) \ge R_{a,2\eps}(P)$. Consider an optimal $\eps$-error protocol for $f$ in the ``a'' model with complexity $R_{a,\eps}(f)$. By the Yao principle, there exists a deterministic protocol for $f$ in the ``a'' model with distributional error at most $\eps$ under $\mu$, and the same complexity. By the union bound, the same protocol computes $P$ with distributional error at most $2\eps$ under $\mu$. The desired inequality $R_{a,\eps}(f) \ge R_{a,2\eps}(P)$ now follows from the choice of $\mu$.
In sum, we have found a function $f$ with $R_{a,\eps}(f) \ge R_{a,2\eps}(P)> R_{b,\eps}(P) \ge R_{b,0}(f)\ge R_{b,\eps}(f)$.

The separation we obtained above is for a functional problem, but not necessarily a \emph{Boolean} one.
We now observe that if the $R_a/R_b$-separation we start with is sufficiently strong, and the number
of output bits of $P$ is not too large, then one can obtain a \emph{Boolean} function with a strong $R_a/R_b$-separation. Assume $f$ has a $k$-bit output, and assume that $g:\{0,1\}^k \to \{0,1\}^{10k}$ is an error-correcting code with constant rate and constant relative distance (which is known to exist). Let $f_1,\ldots,f_{10k}$ be the Boolean functions representing the bits of $g(f(\cdot))$.
Having $\eps$-error protocols for each of these Boolean functions, each of complexity $c$, implies an $\eps$-error
protocol for $f$ of complexity $O(ck)$. Hence for at least one $j$ we have $R_{a,\eps}(f_j)=\Omega(R_{a,\eps}(f)/k)$.
Accordingly, if the original relational $R_a/R_b$-separation was by more than a factor $k$, then we now have a $R_a/R_b$-separation for a Boolean function.

\subsubsection*{Acknowledgments}
We thank Wim van Dam for asking the question that prompted this research, Ben Toner for useful discussions,
and the anonymous CJTCS referees for some comments that improved the presentation of the results.

\bibliographystyle{plain}

\end{document}